\documentclass[preprint]{aastex}

\renewcommand{\deg}{{\ifmmode^\circ\else$^\circ$\fi}}  

\slugcomment{Accepted for publication in {\it Astronomical Journal}} 
 
\input psfig.sty

\shorttitle{KNUDSON, RATNATUNGA, GRIFFITHS}
\shortauthors{GRAVITATIONAL LENS MASS MODELS}

\begin{document}

\title{Investigation of Gravitational Lens Mass Models}

\author{Adam Knudson, Kavan U. Ratnatunga, \& Richard E. Griffiths}

\affil{Physics Dept., Carnegie Mellon University, \\
Pittsburgh, PA 15213 \\
 knudson kavan, \& griffith@astro.phys.cmu.edu}

\begin{abstract}

We have previously reported the discovery of strong gravitational 
lensing by faint elliptical galaxies using the WFPC2 
on HST and here we investigate their potential usefulness
in putting  constraints on lens mass models.  
We compare various ellipsoidal surface mass distributions,
including those with and without a core radius, as well as
models in which the mass distributions are assumed to have the same axis ratio
and orientation
as the galaxy light. We also study models which use a spherical mass
distribution having various profiles, both empirical and following those 
predicted by CDM simulations.  These models also include a gravitational
shear term.  The model parameters and associated
errors have been derived by 2-dimensional analysis of the observed HST
WFPC2 images. The maximum likelihood procedure iteratively converges
simultaneously on the model for the lensing elliptical galaxy and the
lensed image components.  The motivation for this study was to distinguish
between these mass models with this technique.  However,
we find that, despite using the full image data rather than just locations 
and integrated magnitudes, the lenses are fit equally well with several of 
the mass models. Each of the mass models generates a similar configuration
but with a different magnification and cross-sectional area within the caustic,
and both of these latter quantities govern the discovery probability
of lensing in the survey.
These differences contribute to considerable cosmic scatter in any estimate
of the cosmological constant, $\Lambda$ using gravitational lenses.

\end{abstract}
\keywords {cosmology:observations - gravitational lensing - surveys}

\section {INTRODUCTION}
Samples of multi-component gravitational lenses 
(the prototype being the ''Einstein Cross'') have heretofore 
been used to try to constrain the mass model for the 
lensing galaxy \citep[e.g.]{koc95,coh00} and also to constrain 
cosmological models (for a review see \citet{mel99}).
The MDS data have been used to identify 
candidates for such strong gravitational lens systems, based on 
elliptical galaxies.
Such lenses, observed with the angular
resolution and depth
of space based observations, can contribute to the 
study of the mass models and should eventually allow  the compilation
of  new samples
which can be used  to probe cosmology. 
These systems discovered using HST have
critical radii which are mostly sub-arcsecond \citep{rat99}
and have therefore been difficult objects for 
ground-based spectroscopic confirmation \citep[e.g.]{cra96}.
The aim of this paper is to try
to constrain the mass models of the observed lenses.

The deflection of light is proportional to the
gravitational potential gradient at the  locations of the lensed 
images.  The
characteristic separation $\Delta\theta$ of the lensed images depends
on the lensing surface mass density and the distances between the
observer, the lens, and the source object. This allows the
mass-to-light ratio of the lens galaxy to be determined directly if
the lens and source redshifts are known.  The observed lens
configuration is a limited probe of the mass distribution of a
galaxy. Since only a few free parameters can be constrained by the
observations, we need to adopt realistic but simple models for the
projected mass distribution.

\citet[hereafter RGO99]{rat99} adopted the
simple model of a singular isothermal elliptical potential \citep{kor94}
which provides a sufficiently accurate
description of the gravitational lens candidates.  In this paper we
continue this study and investigate six different models of mass
distributions for the lenses. These models incorporate many different
features, including 
  (i) constraining the axis ratio and orientation for the galaxy mass
      distribution to equal the galaxy light distribution,
 (ii) a core radius, 
(iii) a softness parameter on the empirical profile, 
 (iv) N-body simulated CDM profiles \citep{nfw95}, and
  (v) a global shear when required.
The motivation for this study is to examine the structure of
the lens galaxies by identifying the best fit mass model.

We therefore investigate the dependence of the 
magnification of the lens system and the area within the caustic 
over the possible range of surface mass distributions as represented by
different models.

For a given lens configuration, we can derive a range of potentials
that may give an equally acceptable fit to the data.  An understanding
of this range of potential distributions will give us a better
understanding of the range of possible mass distributions for the lens
galaxy. In turn, this will lead to a better understanding of the cosmic
scatter in the parameter estimates.

We discuss the characteristics of the detected sample to understand
the selection effects and the discovery space. With the exception of
our first lens discovery \citep{rat95}, all of
these candidates require confirmation by spectroscopic observations
which are difficult from the ground at these faint magnitudes and
image separations.

\section{THE MODELS FOR THE LENS POTENTIAL}
In the following section we summarize several well known mass models
with a common notation.  The models are also normalized such that if
the mass is spherical and the source centered on the lens, the resulting
Einstein ring will have a common radius for all models.  

For modeling the gravitational lenses in this paper we used the
dimensionless lens equation,
\begin{equation}
\mathbf{y}=\mathbf{x-\alpha\!\left(x\right)},  \label{eq:lens}
\end{equation}
where $\mathbf{y}$ is the source position $\mathbf{x}$ is the impact vector of a light
ray in the lens plane and $\mathbf{\alpha\!\left(x\right)}$ is the deflection angle.
The deflection angle is determined from the (dimensionless) surface mass density $\kappa\!\left(\mathbf{x}\right)$
This is done by first calculating a deflection potential given by
\begin{equation}
\psi\left(\mathbf{x}\right)=\frac{1}{\pi}\int_{\mathrm{I}\!\mathrm{R}^{2}}\kappa\left(\mathbf{x^{\prime}}\right)\ln|\mathbf{x-x^{\prime}}|d^{2}\!x^{\prime}.
\end{equation}
This potential is related to the deflection angle and the surface mass density by
\begin{equation}
\mathbf{\alpha\!\left(x\right)}=\nabla\!\psi\left(\mathbf{x}\right),
\end{equation}
and
\begin{equation}
\nabla^{2}\!\psi\left(\mathbf{x}\right)=2\kappa\!\left(\mathbf{x}\right).
\end{equation}

The models used in this study were chosen so that a variety of key
parameters could be tested. The first model used was the same as that
used by RGO99, which is a singular isothermal elliptical mass
distribution (hereafter SIE).  The normalized two dimensional mass
distribution is given by the following equation 
\begin{equation}
\kappa(x,\varphi)=\frac{\kappa_{0}\sqrt{f}}{2b}
\end{equation}
with 
\begin{equation}
b=x\sqrt{f^2\cos^{2}\!\varphi+sin^{2}\!\varphi}
\end{equation}
where $f$ is the ratio of the minor axis to the major axis and $\kappa_{0}$
is the critical radius (which is the radius of the Einstein ring if the 
lens is spherical and the source centered on galaxy). We will also use $f^{\prime}=\sqrt{1-f^{2}}$ to simplify the following equations.  
This surface mass density results in a deflection angle of
\begin{equation}
\mathbf{\alpha\left(x\right)}=\frac{\kappa_{0}\sqrt{f}}{f^{\prime}}\left[\mathrm{arcsin}\left(f^{\prime}\sin\!\varphi\right)\mathbf{e_{1}}-\mathrm{arcsinh}\left(\frac{f^{\prime}}{f}\cos\!\varphi\right)\mathbf{e_{2}}\right]
\end{equation} 
 A slight
modification to this model is to include a core radius $x_{c}$ giving
\begin{equation}
\kappa(x,\varphi)=\frac{\kappa_{0}\sqrt{f}}{2\sqrt{b^{2}+x_{c}^{2}}},
\end{equation}
Which gives a deflection angle of
\begin{equation}
\alpha_{1}\left(x_{1},x_{2}\right)=\frac{\kappa_{0}\sqrt{f}}{2f^{\prime}}\left(\arg\!S-\arg\!R\right),
\end{equation}
\begin{equation}
\alpha_{2}\left(x_{1},x_{2}\right)=\frac{\kappa_{0}\sqrt{f}}{4f^{\prime}}\ln\frac{Q_{+}}{Q_{-}},
\end{equation}
where
\begin{equation}
R=x_{1}^{2}+f^{4}x_{2}^{2}-f^{\prime2}\left(b^{2}+x_{c}^{2}\right)-2\mathrm{i}f^{2}f^{\prime}\sqrt{b^{2}+x_{c}^{2}}\,x_{2}
\end{equation}
\begin{equation}
S=f^{2}x^{2}-f^{\prime2}x_{c}^{2}-2\mathrm{i}f^{2}f^{\prime}x_{c}x_{2}
\end{equation}
\begin{equation}
Q_{\pm}=\frac{\left(f^{\prime}\sqrt{b^{2}+x_{c}^{2}}\pm x_{1}\right)^{2}+f^{4}x_{2}^{2}}{\left(fx^{2}\pm f^{\prime}x_{c}x_{1}\right)^{2}+f^{\prime2}x_{c}^{2}x_{2}^{2}}
\end{equation}
We will refer to this model as the non-singular isothermal
ellipsoid (hereafter NIE).  These models can both be found in 
\citet[hereafter KSB94]{kor94} with the exception 
that our distributions are rotated 90\deg with respect to KSB94.  Another
variation on the SIE model was to constrain the orientation and
axis ratio of the model mass to coincide with that of the light
giving a model in which the shape of the mass traces the light (hereafter SML).
One final variation of the SIE models was to constrain only the orientation
of the model mass so that it was the same as that of the light (hereafter SIG). 
In order to fit the observed images these constant mass to light ratio
models also required that a shear term of the form $-\left(\gamma/2\right)x^{2}\cos (\varphi-\theta_{\gamma})$
be added to the potential.  The two remaining models both
assume a spherical mass distribution and so they also require a shear
term to break the spherical symmetry.  The first of these is a
singular spherical mass distribution with a softness parameter (hereafter SSS)
giving the mass distribution, 
\begin{equation}
\kappa(x)=\left(\frac{\beta}{2}\right)\left(\frac{\kappa_{0}}{x}\right)^{(2-\beta)}
\end{equation}
where $\beta=1$ corresponds
to the singular isothermal sphere and $\beta=0$ describes a point mass
\citep{wit95}.
The deflection angle for the SSS model is
\begin{equation}
\mathbf{\alpha}\!\left(x\right)=\kappa_{0}^{\left(2-\beta\right)}x^{\left(\beta-1\right)}
\end{equation}
 The other spherical model uses a density
profile derived from N-body simulations by \citet{nfw95}
(hereafter NFW).  This NFW mass distribution is given by.
\begin{equation}
\kappa(x)=\frac{\kappa_{0}^{2}f(x)}{2\left(\ln\!\left(1/2\right)+1\right)\left(x^{2}-1\right)}
\end{equation}
where
\begin{equation}
f(x)=\cases{1-\frac{2\kappa_{0}}{\sqrt{x^{2}-\kappa_{0}^{2}}}\arctan\sqrt{\frac{x-\kappa_{0}}{x+\kappa_{0}}} & 
$\left(x>\kappa_{0}\right)$ \cr
1-\frac{2\kappa_{0}}{\sqrt{\kappa_{0}^{2}-x^{2}}}\mbox{arctanh}\sqrt{\frac{\kappa_{0}-x}{\kappa_{0}+x}} & 
$\left(x<\kappa_{0}\right)$ \cr
0 & $\left(x=\kappa_{0}\right)$ \cr }
\end{equation}
giving
\begin{equation}
\mathbf{\alpha}\left(x\right)=\frac{\kappa_{0}^{2}g\!\left(x\right)}{x\left(\ln\!\left(1/2\right)+1\right)}
\end{equation}
with
\begin{equation}
g\!\left(x\right)=\ln\!\frac{x}{2\kappa_{0}}+\cases{\frac{2\kappa_{0}}{\sqrt{x^{2}-\kappa_{0}^{2}}}
\arctan\sqrt{\frac{x-\kappa_{0}}{x+\kappa_{0}}} & $\left(x>\kappa_{0}\right)$ \cr
\frac{2\kappa_{0}}{\sqrt{\kappa_{0}^{2}-x^{2}}}\mbox{arctanh}\sqrt{\frac{\kappa_{0}-x}{\kappa_{0}+x}} & 
$\left(x<\kappa_{0}\right)$ \cr
1 & $\left(x=\kappa_{0}\right)$. \cr}
\end{equation}

\begin{figure}[h]
 \centerline{\psfig{file=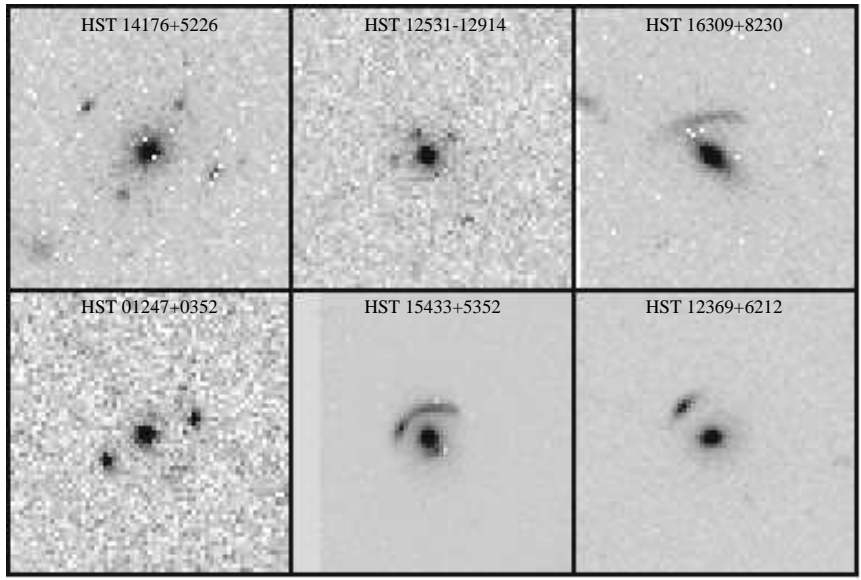,height=3.0in}}
 \caption{WFPC2 F606W images of the six lenses used in this study. Each image is 6 \farcs 4 x 6 \farcs 4}
 \label{fig:lens}
\end{figure}

We selected the best six of the ``Top Ten'' MDS gravitational lens candidates
(RGO99) and fitted them with these six models.  These six lenses can 
be seen in Figure \ref{fig:lens}.  Of the four that were not used,
two were too faint, one was too close to the CCD edge where the PSF
changes rapidly and one was found in a small group.  These difficulties 
made these four lenses undesirable for comparison of the  mass models.  
We used the same software procedure as used in RGO99 with the required 
modification to the potential.

We also attempted to model three previously reported lenses with HST images:  
PG1115+080 \citep{kri93}, MG0414+0534 \citep{fal97}, and B1608+656
\citep{mye95}. In all of these cases the lensed images are brighter than
the lensing galaxy.  The relative fluxes of the lensed components are well
constrained by the high signal to noise ratio of these images but are 
poorly fit by simple lens models. This is a well known problem that can 
only be solved by using more complicated mass models which include 
substructure in the lens galaxy \citep{mao98}.  Since the fit and residuals
were dominated by the poor fit to the lensed component fluxes they were 
also undesirable for comparison of mass models.

Given a set of model parameters, we generate 2-dimensional images for
the elliptical and the source galaxies.  The expected configuration of
the lensed images is ray-traced.  To improve
the accuracy of this calculation, the image is evaluated
using a pixel size smaller than the real one, such that the image
fills a 64 pixel square array. The elliptical lens and the lensed
source images are then convolved with the adopted HST WFPC2 point
spread function from TinyTim \citep{kri00}.  The convolved image is
spatially integrated to the WFC pixel size of 0\farcs1 and is then
compared with the observed galaxy image. The likelihood function is
defined as the sum over all pixels of the logarithm of the probability of
the observed value, assuming a Gaussian error distribution
with respect to the model.  This is proportional to a weighted $\chi^2$,
and is then minimized using a quasi-newtonian method.  The rms errors
are estimated from the covariance matrix which is derived by inverting
the Hessian at the maximum of the likelihood function.  This method is 
sensitive not only to the location and magnitude of the lensed components 
but also to their shape and distortion. Once a best fit was found for 
each model we calculated the area within the caustic. The caustic is 
found by first calculating the image distortion which can be expressed 
by the Jacobian matrix: 
\begin{equation}
A\left(\mathbf{x}\right)=\frac{\partial\mathbf{y}}{\partial\mathbf{x}}
\end{equation}
Next the critical curves are found by solving $\det\mathbf{A}=0$. This
solution is then inserted into the lens equation (eq. \ref{eq:lens}),
giving the caustics.  Most models have two caustics. However, some
models (SIE and its variants as well  as SSS for certain parameters)
have only one.  In these cases we calculate a cut which is a curve
surrounding the region where multiple images exist, but which is not
within the caustic \citep{kov87}.  The cut becomes a caustic if the
singularity in the surface mass distribution is removed.  The cut and
caustic were computed analytically for SIE, SML, SIG and NIE using
equations (slightly modified to  include a shear term) found in KSB94. 
 The cut and caustic for the SSS was also  computed analytically, while
the NFW's caustics had to be determined numerically.   The details of
the cut and caustic equations can be found in the Appendix.  The areas
within the caustics and/or cut were then integrated numerically.
 
All six models used the standard parameters for modeling the light
distribution of the lens galaxy: local sky, location (x,y), total
magnitude, half-light radius, axis ratio and position angle. In
addition, all models contained parameters describing the properties
of the lensed source:  x and y offset with respect to the lens 
galaxy, magnitude, half-light radius, axis ratio and orientation.  
Parameters describing the mass include the critical radius
plus the following model dependent 
parameters:  the axis ratio and orientation of the lens mass for the 
SIE and NIE models, a core radius for the NIE, a 
softness parameter for the SSS, and two parameters to describe the 
global shear (magnitude and direction) for the SSS, NFW, SML, and SIG.
For all models the centroid of the gravitational potential of the 
lens was assumed to be the same as that of the light of the lensing galaxy.
For reference the six models and a brief explanation of each are listed
in Table \ref{tab:mass}.

\begin{table}[h]
 \begin{center}
 \caption{Explanation of mass models.\label{tab:mass}}
 \begin{tabular}{lp{1.0in}p{4.0in}}
   \tableline
   \tableline
   {\bf Model} & {\bf Number of Parameters} & {\bf Description}\\
   \tableline
    SIE & 15 & Singular Isothermal Ellipsoid.\\
    SIG & 16 & Same as SIE with mass orientation the same as the light. Requires shear.\\
    SML & 15 & Same as SIE with mass orientation \& axis ratio the same as the light. Requires shear.\\
    NIE & 16 & Isothermal Ellipsoid with a core radius (non-singular).\\
    SSS & 16 & Singular Spherical model with a Softness parameter.  Requires shear.\\
    NFW & 15 & Spherical model from CDM simulations.  Requires shear.\\
   \tableline
 \end{tabular}
 \end{center}
\end{table}

Unlike the fully automated Disk-plus-Bulge decomposition in the Medium
Deep Survey pipeline \citep{rat99b}, convergence of the lens model
requires an initial guess in which the lensed images at least overlap
the observed components. Since the number of discovered lenses is
limited, we have so far worked out the initial guess by trial and error.
We have also simplified the derivation of the initial guess by using a
web based simulator. This form based cgi-driver, which includes all the 
model parameters, has been written in f77 using the same subroutines 
used in the lens analysis.

\section{ANALYSIS}

In most cases several models fit a given lens system equally well and rarely was
any model strongly ruled out.  This can be seen in Table \ref{tab:mle} which shows
the maximum likelihood for all lenses and passbands.  For each lens, the likelihood
values for each passband of a given model were summed.  The best fit
model is then just the model for which this sum is the smallest.  Any
other model with a summed likelihood  which differs from the best by
less than 5 times the number of passbands  was considered to be
statistically equivalent to the best model, although even models  with
much larger differences still gave reasonable fits to the lens systems.

\begin{table}[h]
 \begin{center}
 \caption{Maximum likelihood for each lens system and passband.\label{tab:mle}}
 \footnotesize
 \begin{tabular}{ccccccc}
  \tableline\tableline
  \multicolumn{3}{c} {\bf HST 01247+0352} &  \multicolumn{4}{c} {\bf HST 12369+6212} \\
  \tableline
  {\bf Model} & {\bf F606W} & {\bf F814W}  & {\bf Model} & {\bf F606W} & {\bf F814W} &{\bf F450W}\\
  \tableline {\bf SIG} & 2536.7 & 2682.3 & {\bf SSS} & 1924.2 & 3115.1 & 1613.9\\
  {\bf SSS} & 2542.8 & 2682.2 & {\bf SIG} & 1927.0 & 3115.9 & 1610.9\\
  NFW & 2560.8 & 2712.3 & {\bf NFW} & 1927.1 & 3117.8 & 1610.3\\
  SIE & 2561.8 & 2702.5 & {\bf NIE} & 1933.6 & 3105.6 & 1623.4\\
  NIE & 2562.4 & 2703.2 & {\bf SIE} & 1934.2 & 3106.4 & 1623.5\\
  SML & 2564.8 & 2704.8 & SML & 1939.5 & 3121.9 & 1609.3\\
  {\bf \# of pixels} & 1902 & 1917 & {\bf \# of pixels} & 1281 & 1594 & 1168 \\
  \tableline
  \multicolumn{3}{c} {\bf HST 12531-2914} & \multicolumn{4}{c} {\bf HST 15433+5352} \\
  \tableline {\bf Model} & {\bf F606W} & {\bf F814W}  & {\bf Model} & {\bf F606W} & {\bf F814W} &{\bf F450W}\\
  \tableline {\bf SIG} & 1626.5 & 2057.5 & {\bf SSS} & 3286.7 & 2913.9 & 1956.3\\
  {\bf NFW} & 1627.8 & 2054.1 & NFW & 3321.2 & 2952.4 & 1950.8\\
  {\bf SSS} & 1628.0 & 2055.5 & SIG & 3323.0 & 2944.8 & 1925.2\\
  {\bf SML} & 1628.5 & 2054.8 & SML & 3327.3 & 2946.4 & 1934.5\\
  {\bf NIE} & 1633.2 & 2053.0 & NIE & 3399.4 & 2975.4 & 1966.4\\
  {\bf SIE} & 1636.7 & 2052.7 & SIE & 3399.6 & 2970.8 & 1971.1\\
  {\bf \# of pixels} & 1168 & 1416 & {\bf \# of pixels} & 1758 & 1780 & 1298 \\
  \tableline \multicolumn{3}{c} {\bf HST 14176+5226} & \multicolumn{4}{c} {\bf HST 16309+8230} \\
  \tableline {\bf Model} & {\bf F606W} & {\bf F814W}  & {\bf Model} & {\bf F606W} & {\bf F814W} &{\bf F450W}\\
  \tableline {\bf SIG} & 3403.8 & 4477.2 & {\bf NFW} & 2039.5 & 2544.8 & 1864.8\\
  {\bf NIE} & 3410.3 & 4466.5 & {\bf SIE} & 2040.9 & 2546.1 & 1867.2\\
  {\bf SIE} & 3410.4 & 4471.0 & {\bf NIE} & 2043.1 & 2542.7 & 1864.4\\
  SML & 3415.3 & 4479.6 & {\bf SSS} & 2045.2 & 2546.1 & 1868.9\\
  SSS & 3425.1 & 4487.4 & {\bf SIG} & 2045.3 & 2547.7 & 1865.3\\
  NFW & 3447.2 & 4492.3 & {\bf SML} & 2046.3 & 2548.4 & 1866.9\\
  {\bf \# of pixels} & 2453 & 3015 & {\bf \# of pixels} & 1528 & 1781 & 1438 \\
  \tableline
 \end{tabular}
 \tablecomments{Bold model names are the best fit models}
 \end{center}
\end{table}

\begin{figure}[h]
 \centerline{\psfig{file=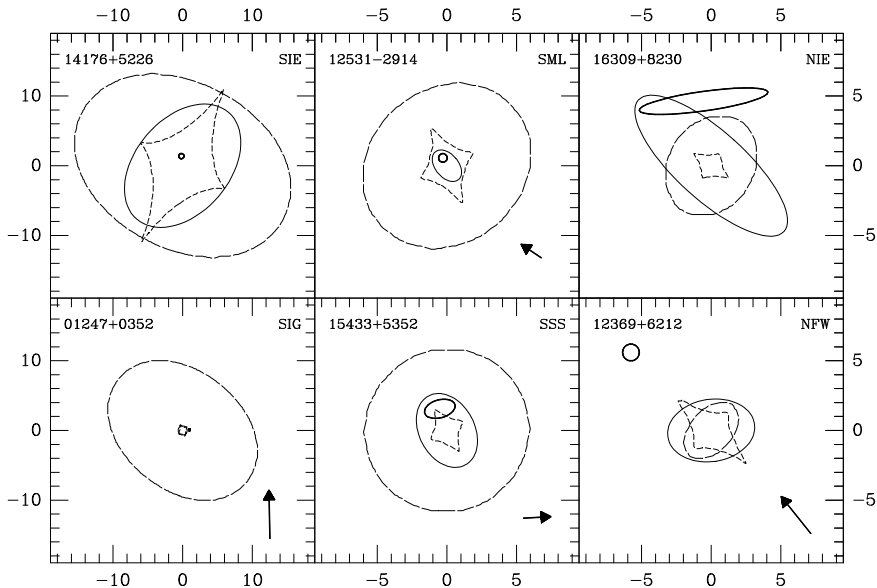,height=3.25in,angle=-90}}
 \caption{A sample of cut and caustic images in F606W.  Short dashes show the interior caustic, long dashes are the outer caustic (or cut), and a thin/thick solid line represents the ellipse with the half light radius of the lens/source galaxy. The arrow shows the relative magnitude and direction of the shear where applicable.}
 \label{fig:modsamp}
\end{figure}

The first lens (HST 14176+5226) was our best four-group image.  Several models fit
equally well including the SIE, NIE, and SIG.  The SSS, NFW and SML models
do not fit as well but are not strongly ruled out.  The only other four group image was HST 12531-2914.  
All models gave indistinguishable results for this lens.  The lens HST 01247+0352, which consists
of a nearly spherical lens galaxy and two images of the source galaxy, was fit
equally well with SSS and SIG while the remaining models were again not strongly
ruled out.  The other two-component lens image was
HST 15433+5352 which was made up of an elliptical lens galaxy, a large arc, and a small
source image opposite the arc.  The SSS model was favored over the other models
for this system: this is the only lens that favored just one model.  The
remaining two lenses modeled were both cases of a lens galaxy and a
single arc.  The first of these, HST 12369+6212,  was from the Hubble Deep Field and was
first reported by \citet{hog96}.   It was modeled  well by SSS, SIG, SIE, NIE, and NFW.  The
final lens, HST 16309+8230, showed no preference for any model.    
Thus many models give adequate fits to the same lenses, introducing a
degeneracy in the parameters describing the mass of the lens systems. 
An example of the resulting caustic image for each system is shown in
Figure \ref{fig:modsamp} and Table  \ref{tab:sample} gives some of the
parameters derived from  modeling the systems in Figure \ref{fig:modsamp}.  The 
full list of all parameters for all of the lens system can be found at
the web site http://mds.phys.cmu.edu/lenses.

\begin{table}[h]
 \begin{center}
   \caption{A sample of basic parameters in F606W.\label{tab:sample}}
 \begin{tabular}{p{2.25in}p{36pt}p{36pt}p{36pt}p{36pt}p{36pt}p{36pt}}
   \tableline
   \tableline {\bf Parameter}  &  {\bf  \small 14176 $+$5226} & {\bf \small 01247 $+$0352} & {\bf \small 12369 $+$6212} &
       {\bf \small 16309 $+$8230} & {\bf \small 12531 $-$2914} & {\bf \small 15433 $+$5352}\\
   \tableline {\bf Model} & SIE & SIG & NFW & NIE & SML & SSS\\  
   {\bf \footnotesize Lens Magnitude} & 21.84 & 23.95 & 23.96 & 21.61 & 23.89 & 21.86\\
   {\bf \footnotesize Lens half-light radius ($arcsec$)}  & 1.012 & 0.065 & 0.315 & 0.705 & 0.130 & 0.281\\
   {\bf \footnotesize Lens position angle ($deg$)} & $-$40.62 & $-$38.67 & $-$81.42 & 48.01 & 40.05 & 27.90\\
   {\bf \footnotesize Lens axis ratio}  & 0.67 & 1.00 & 0.71 & 0.35 & 0.64 & 0.71\\
   {\bf \footnotesize Source x offset ($arcsec$)}  & $-$0.019 & 0.097 & $-$0.577 & $-$0.055 & $-$0.029 & $-$0.051\\
   {\bf \footnotesize Source y offset ($arcsec$)}  & 0.138 & 0.015 & 0.560 & 0.463 & 0.056 & 0.155\\
   {\bf \footnotesize Critical Radius ($arcsec$)}  & 1.496 & 1.085 & 0.440 & 0.414 & 0.595 & 0.577\\
   {\bf \footnotesize Mass Axis Ratio}  & 0.37 & 0.37 & 1.00: & 0.59 & 0.64: & 1.00:\\
   {\bf \footnotesize Angle of Mass wrt Light ($deg$)}  & 12.20 & 0.00: & 0.00: & 8.54 & 0.00: & 0.00:\\
   {\bf \footnotesize Source half-light radius ($arcsec$)}  & 0.038 & 0.010 & 0.060 & 0.467 & 0.031 & 0.113\\
   {\bf \footnotesize Source position angle ($deg$)}  & 0.00: & 0.00: & 0.00: & $-$81.93 & 0.00: & $-$74.60\\
   {\bf \footnotesize Source axis ratio} & 1.00: & 1.00: & 1.00: & 0.15 & 1.00: & 0.57\\
   {\bf \footnotesize Softness parameter}  & 1.00: & 1.00: & NA & 1.00: & 1.00: & 0.92\\
   {\bf \footnotesize Core radius}  & 0.000: & 0.000: & 0.000: & 0.005 & 0.000: & 0.000:\\
   {\bf \footnotesize Shear magnitude}  & 0.000: & 0.300 & 0.295 & 0.000: & 0.128 & 0.150\\
   {\bf \footnotesize Shear direction ($deg$)}  & 0.00: & $-$88.87 & $-$51.34 & 0.00: & $-$33.40 & 3.04\\
   \tableline
 \end{tabular}
 \tablecomments{The `:' indicates parameters that were not fitted, but fixed at the values shown.}
 \end{center}
\end{table}

The fact that four of the six models use a shear term allows us to investigate
external shear for the lenses.  Most of	the lens systems required small
values of shear roughly consistent with cosmic shear due to large scale
structure \citep{cas99}.  The only exceptions to this are HST 12369+6212
and HST 12531-2914 both of which require shears that are large compared
to external shears expected from large-scale structure.  In the case of 
12531-2914 we find that a shear of $\approx0.14$ is required to fit the
observed configuration.  This value is slightly less than the value of 
$\approx0.20$ found by \citet{wit97}.
This shear could be due to a nearby galaxy or by dropping the requirement that the mass of the
lens be aligned with the light.  Since there does not appear to be a
galaxy at the require angle of $\approx-26\deg$ we agree with the findings of \citet{wit97}
who concluded that the shear was due to a misalignment of mass and light.
In fact, the lens system was well fit by allowing the mass to be rotated 
by $\approx-15\deg$.  HST 12369+6212 was found to have a shear term of
$\approx0.25$ at a direction of $\approx-54\deg$.  Two nearby objects
(12:36:57.10+62:12:26.7 and 12:36:56.31+62:12:10.2) can be found in this 
direction which could explain this large shear.  Although this lens was also well
fit by allowing the position angle of the mass to vary the fitted angle was very 
large ($\approx-68\deg$) making it a less attractive model.

\begin{figure}[h]
 \centerline{\psfig{file=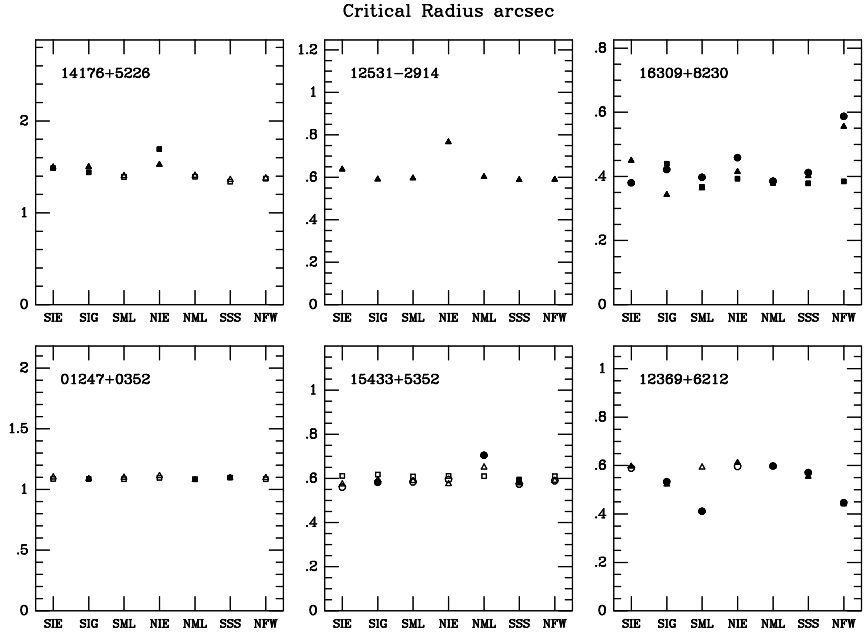,height=3.25in,angle=-90}}
 \caption{Critical radius for each lens system and model. Square:  F814W, triangle: F606W, and circle: F450W.  Filled symbols represent the best fit models }
 \label{fig:crit}
\end{figure}

We next compare in Figure \ref{fig:crit}  the fractional change in the
critical radius as a function of the models fitted to each of the six
gravitational lens systems. Since the critical radius is determined by the
redshifts of the lens and source and the velocity dispersion of the
lens we expect that it should not vary significantly between models.
We find that the critical radius change is
4.3\% rms. We note that the SSS and NFW model fits have a slightly
smaller critical radius than SIE. This result confirms the theoretical
expectation that if we were able to measure the velocity dispersion with
an accuracy of a few km/s, then the cosmological constant could be
directly estimated \citep{ime97}.

\begin{figure}[h]
\centerline{\psfig{file=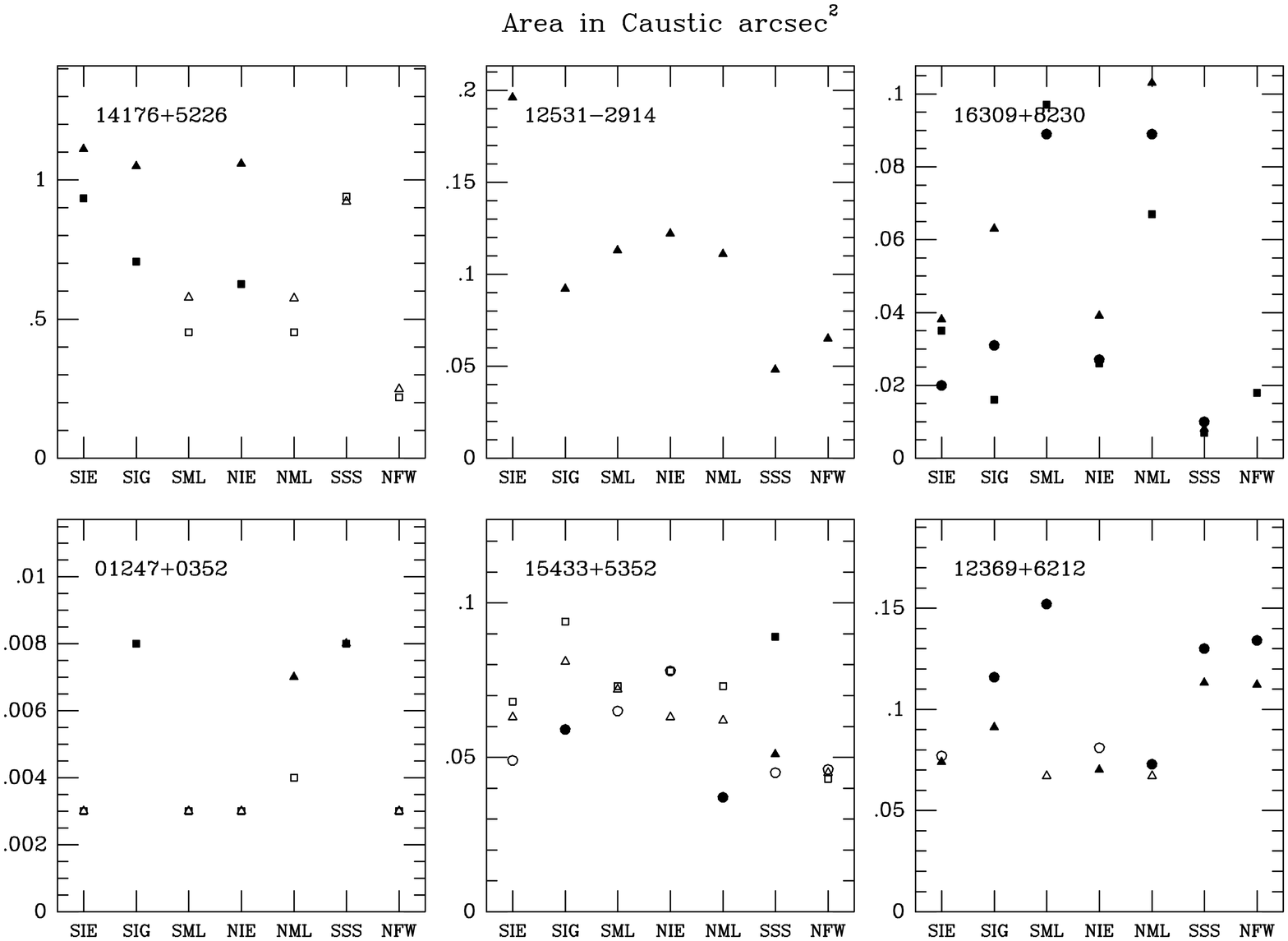,height=3.25in,angle=-90}}
 \caption{Area within interior caustic for each lens system and model. Square:  F814W, triangle: F606W, and circle: F450W.  Filled symbols represent the best fit models }
 \label{fig:caus}
\end{figure}

The surface mass density distribution and the resulting potential
gradient change lead to a small caustic centered on each lensing galaxy.
For different potential models fitted to the same observed lens configuration, we see
that the fractional change in the area within the caustic is much
larger than that expected from the small change in critical radius.  We
note that the area within the interior caustic changes by 25\% rms between models 
of the same lens configuration (Figure \ref{fig:caus}).
The SSS and NFW models have a slightly smaller area within the
caustic for strong gravitational lens cases.  This result agrees well with \cite{bla87} 
which showed that shallower mass profiles (such as NFW and SSS with $\alpha$ less than 1.0) 
have smaller cross sections.  This result indicates
that the range of area within the caustic is a possible contributor to 
errors in the estimation of the number of detectable lenses expected
from a survey. 

\begin{figure}[h]
\centerline{\psfig{file=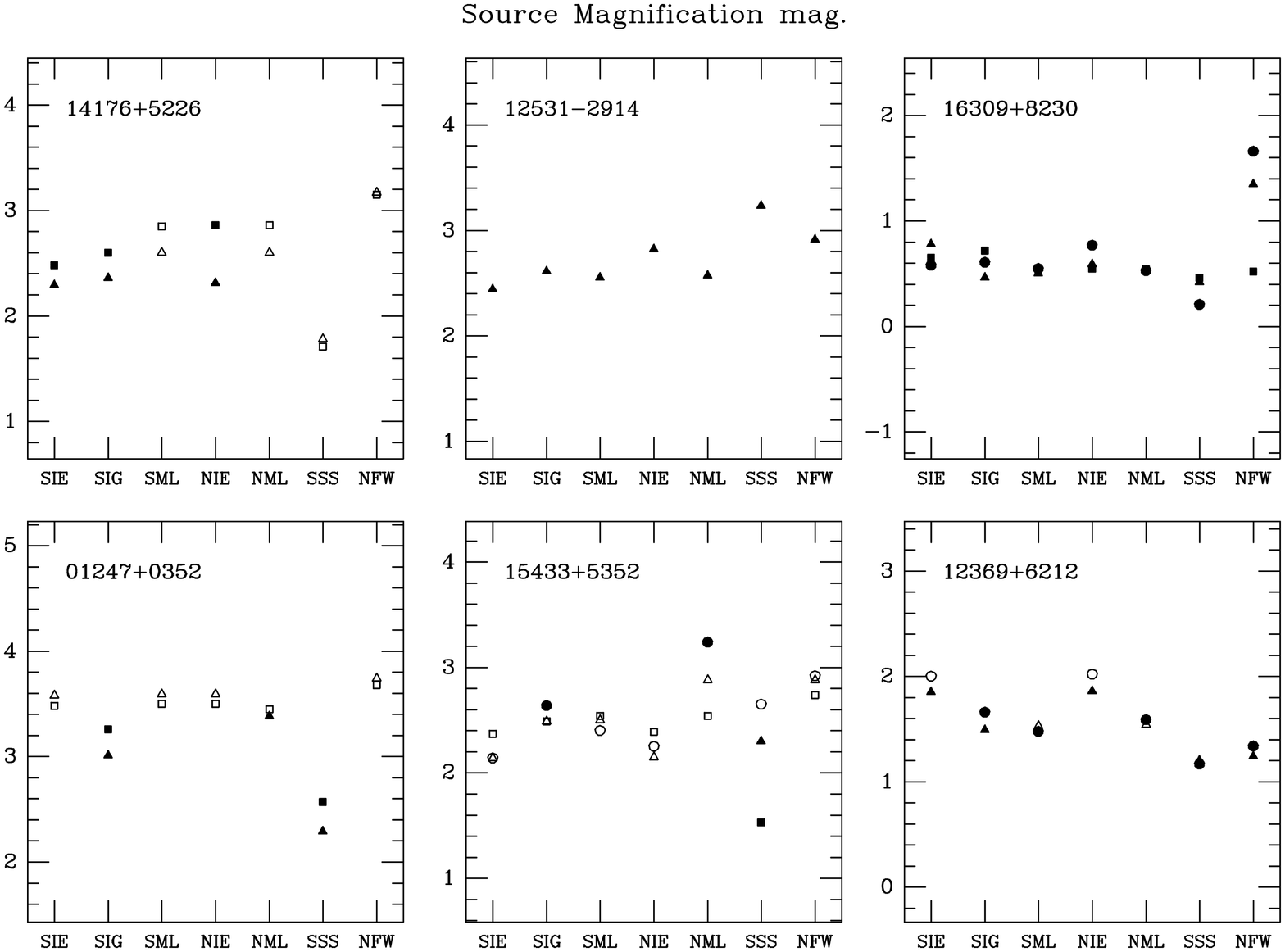,height=3.25in,angle=-90}}
 \caption{Magnitude for each lens system and model. Square:  F814W, triangle: F606W, and circle: F450W.  Filled symbols represent the best fit models }
 \label{fig:mag}
\end{figure}

In a magnitude limited sample the discovery probability depends also
on magnification. Figure \ref{fig:mag} shows the variation in magnification
for the various models.  Since each model fits the observed magnitude, a lens
model which has higher magnification implies a fainter source galaxy.
Since there are more galaxies at fainter magnitudes, the discovery
probability increases. However, we note (again in agreement with \cite{bla87}) 
that the models with larger
magnification tend to have smaller caustics.  Thus, these two effects 
oppose each other since a smaller caustic decreases the discovery 
probability.  

\begin{figure}
\centerline{\psfig{file=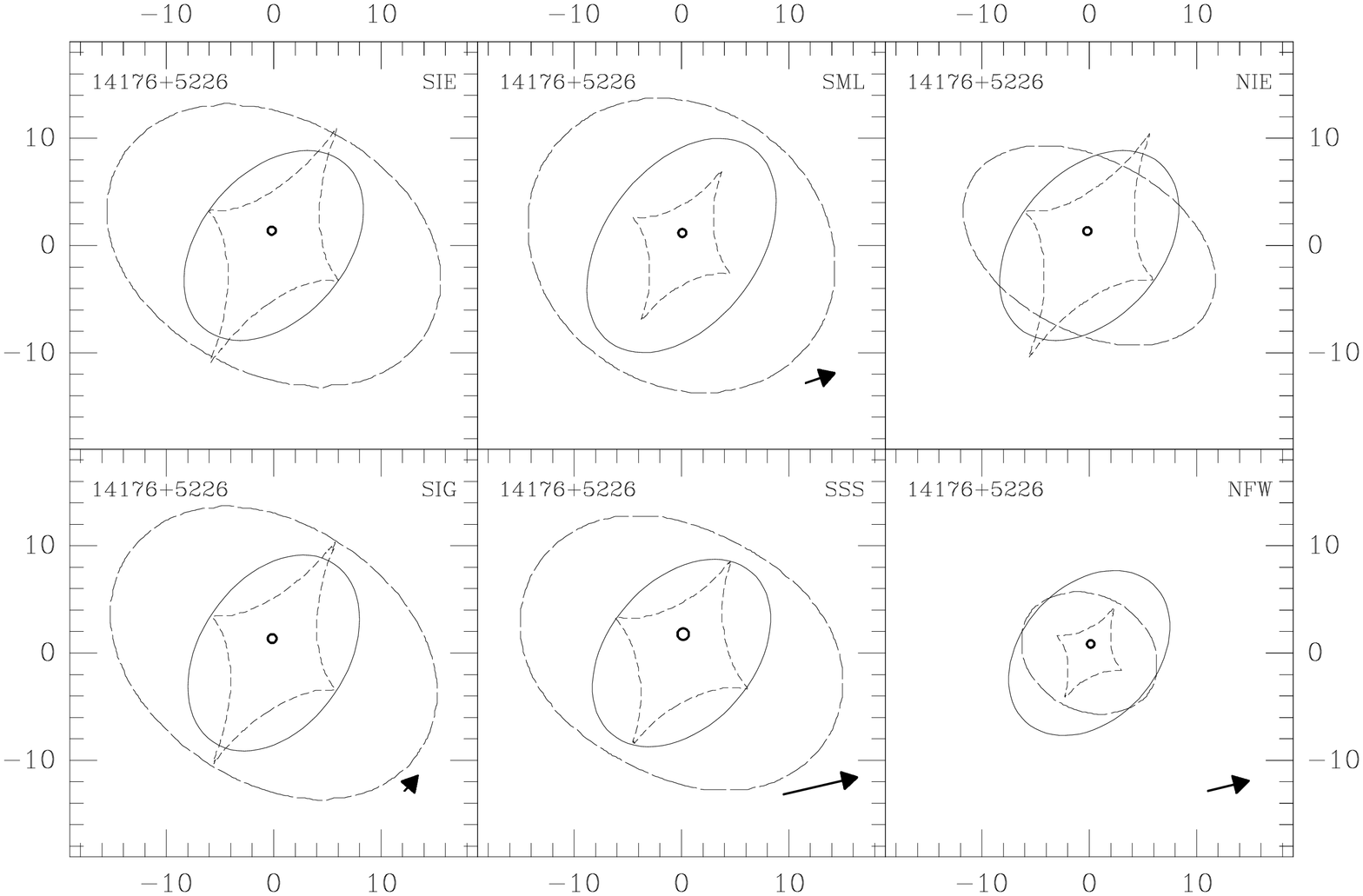,height=3.25in,angle=-90}}
 \caption{Caustic and Cut images for HST 14176+5226 in F606W.  Short dashes show the interior caustic, long dashes are the outer caustic (or cut), and a thin/thick solid line represents the ellipse with the half light radius of the lens/source galaxy. The arrow shows the relative magnitude and direction of the shear where applicable.}
 \label{fig:u26x8f}
\end{figure}

Figure \ref{fig:u26x8f} graphically shows an example of how parameters
can vary from model to model for a given lens system, in this case
HST 14176+5226 which is the only spectroscopically confirmed lens.  
In particular, the size of the interior caustic changes significantly 
between models.  Table \ref{tab:u26x8samp} gives the values  
of many of the parameters in each model for this lens system.

\begin{table}
  \begin{center}
    \caption{A sample of key parameters for HST 14176+5226 F606W.\label{tab:u26x8samp}}
  \begin{tabular}{lcccccc}
    \tableline
    \tableline {\bf Parameter} & {\bf SIE} & {\bf SIG} & {\bf SML} &
           {\bf NIE} & {\bf SSS} & {\bf NFW}\\
    \tableline {\bf Mass Axis Ratio} & 0.37 & 0.38 & 0.64 & 0.39 & 1.00: & 1.00:\\
    {\bf Angle of Mass wrt Light} ($deg$) & 12.20 & 0.00: & 0.00: & 12.32 & 0.00 & 0.00\\
    {\bf Shear Magnitude} & 0.00: & 0.05 & 0.10 & 0.00: & 0.31 & 0.15\\
    {\bf Shear Direction} ($deg$) & 0.00: & 49.16 & 18.39 & 0.00: & 12.89 & 14.04\\
    {\bf Critical Radius} ($arcsec$) & 1.496 & 1.498 & 1.401 & 1.522 & 1.359 & 1.374\\
    {\bf Source Magnitude} ($mag$) & 26.16 & 26.14 & 26.41 & 26.20 & 25.52 & 26.96\\
    {\bf Magnification} ($mag$)& 2.29 & 2.36 & 2.60 & 2.31 & 1.78 & 3.17\\
    {\bf Caustic Area} ($arcsec^2$) & 1.11 & 1.05 & 0.58 & 1.06 & 0.92 & 0.25\\
    \tableline 
  \end{tabular}
 \tablecomments{The `:' indicates parameters that were not fitted, but fixed at the values shown.}
 \end{center}
\end{table} 

\section{CONCLUSIONS}

We have investigated different potentials and found that it is
difficult in most cases to distinguish between them.

We observe that the critical radius is fairly constant for a 
particular lens system, independent of the model used.  The critical
radius is determined by the distances to the  lens and source galaxy,
which are measurable observables, and the integrated mass. Since we see
only small variations across the various models, this implies that the
integrated mass is not significantly different from model to model.

However, even models that lead to very similar lensed image configurations
have very different caustics.
Since the area within the caustic and the magnification determine the
creation and discovery of strong  lenses, these differences, along with
the faint galaxy count distribution,  change the expected number of
detected lenses.  Any attempt to determine the cosmological
constant from such numbers must take these factors into 
consideration.  

We note that samples of radio gravitational lens candidates are discovered by
detecting a compact multiple component image structure with identical
spectral  characteristics. This sample, although well defined,
discovers mainly lenses of QSO sources, and otherwise depends on the
radio characteristics of  faint galaxies.  On the other hand optical
gravitational lens candidates  detected without spectral observations
need to be based on the likelihood of the configuration compared with
random clustering of galaxies.  A single faint blue galaxy in proximity 
to a red elliptical galaxy is not a reliable lens candidate, since such
 configuration very frequently happens by chance.  However, four
galaxies in a cross pattern around an elliptical galaxy is a reliable
lens candidate  since the chance of this occurring at random is very
 small. In the process of this study we examined several radio lenses
with HST data and found that in most cases, we would not have
identified the lens based only on the WFPC2 image. Likewise, many of the
lenses in our sample, in particular HST 14176+5226 and HST 12531-2914,
were not detected by observations made on the VLA. These two methods
(radio and optical) seem to find different lens candidate samples which
complement each other, allowing a better  understanding of  selection
effects.

The discovery probability for optically discovered lenses is affected 
by the symmetry of the configuration, the contrast of source color 
against that of the lens galaxy,  and the critical radius.  
If a distant galaxy happens by chance to fall within the caustic of a
foreground galaxy we observe  four images of the source known as a
`strong' gravitational lens.  If these images are about one magnitude 
brighter than the object detection threshold, with a critical radius 
larger than about 0.4 arcsecond, then we are confident that we would 
discover the lens with high reliability. As the intrinsic location of the source
passes outside the caustic, three of the multiple images merge into one
and the fourth image on the opposite side becomes much fainter, and may not
be visible. Between the caustic and either the outer caustic or cut two
images are formed. This is also the special case of perfect spherical
symmetry and no external shear when the central caustic vanishes.  As the
source location nears the second caustic (or the cut) one image will become 
fainter making only one image (highly distorted) visible. We do not
expect to discover these lens systems unless the source galaxy is
extended with a half light radius larger than about 0.1 arc seconds. We
refer to these cases as `mild' lensing. For instance if the source 
galaxy is intrinsically extended and almost spherical,
then it would be seen to form a prominent arc oriented with the
curvature centered on the lensing galaxy. This makes a probable lens. If
however the original was  extended but with a large ellipticity, then it
would look like a nearby galaxy with random orientation.

The variety of possible lens configurations is numerous.  
We can investigate the discovery 
probability using the fully interactive simulator which we have
made available publicly on the web at http://mds.phys.cmu.edu/lenses.
This interactive tool can not only provide an initial guess to
match a new lens configuration but can also allow us to explore the multi-dimensional
parameter space to investigate the probability of discovery. 
We urge the reader to visit this web site and try it. All models, model parameters,
and cosmological parameters can be selected.  The source location relative to the
caustic can be input via a line diagram of the caustic and cut. A 
color image is created of the lensed source and lensing galaxy.

This paper is based on observations with the NASA/ESA Hubble Space
Telescope, obtained at the Space Telescope Science Institute, which is
operated by the Association of Universities for Research in Astronomy,
Inc., under NASA contract NAS5-26555.  The HST Archival research was funded
by STScI grant GO8384.

\appendix
\section{APPENDIX}
The general method for finding the caustic for a lens was discussed
above. Here we give more specific details on finding the caustic for 
the SSS model and a small correction that was need for the NIE.  The
caustic equations for SIE can be found in KSB94.

For NIE $\det\mathbf{A}=0$ is equivalent to
\begin{equation}
8c_{3}\kappa^{3}+4c_{2}\kappa{2}+2c_{1}\kappa+c_{0}=0
\end{equation}
as in KSB94 with the following correction to the coefficients:
\begin{equation}
c_{3}=\left(x_{c}f^{3/2}-x_{c}^{2}-x_{c}^{2}f^{2}\right)\Delta^{2}+x_{c}^{2}f^{2}
\end{equation}
\begin{equation}
c_{2}=\left(-2x_{c}f^{3/2}+x_{c}^{2}+x_{c}^{2}f^{2}\right)\Delta^{2}-x_{c}^{2}f^{2}
\end{equation}
\begin{equation}
c_{1}=2x_{c}f^{3/2}\Delta^{2}-f^{3}
\end{equation}
\begin{equation}
c_{0}=f^{3}.
\end{equation}
The resulting critical curve is then used in the lens equation to give the caustic.
For SSS solving $\det\mathbf{A}=0$ gives the critical curve
\begin{equation}
x=\kappa_{0}\left[\frac{P-\sqrt{P^{2}-4\left(\beta-1\right)\left(1-\gamma^{2}\right)}}{2\left(\beta-1\right)}\right]^
{\frac{1}{\beta-2}}
\end{equation}
where
\begin{equation}
P=\beta-\gamma\left(\beta-2\right)\cos\left(2\varphi-2\theta_{\gamma}\right)
\end{equation}
which gives the caustic
\begin{equation}
y_{1}=x\cos\varphi-\kappa_{0}^{\left(2-\beta\right)}x^{\beta-1}
+x\gamma\left[\cos\phi\cos\left(2\theta_{\gamma}\right)+\sin\phi\sin\left(2\theta_{\gamma}\right)\right]
\end{equation}
\begin{equation}
y_{2}=x\sin\varphi-\kappa_{0}^{\left(2-\beta\right)}x^{\beta-1}
+x\gamma\left[\cos\phi\sin\left(2\theta_{\gamma}\right)-\sin\phi\cos\left(2\theta_{\gamma}\right)\right].
\end{equation}
The caustic for the NFW model was found by numerically solving the Determinate of $\mathbf A$ and putting the
resulting critical curve back into the lens equation.

\end{document}